\documentclass[runningheads]{llncs}
\usepackage[T1]{fontenc}
\usepackage{graphicx}
\usepackage{url}
\usepackage{subcaption} 
\usepackage{soul}
\usepackage{comment}
\usepackage{amsmath}
\usepackage{multirow}

\begin{document}
\mainmatter              
\title{IntellBot: Retrieval Augmented LLM Chatbot for Cyber Threat Knowledge Delivery}
\titlerunning{IntellBot}  
%
\author{
Dincy R. Arikkat\inst{1} \and 
Abhinav M.\inst{1} \and 
Navya Binu\inst{1} \and 
Parvathi M.\inst{1} \and 
Navya Biju\inst{1} \and 
K. S. Arunima\inst{1} \and 
Vinod P.\inst{2,1} \and
Rafidha Rehiman K. A.\inst{1} \and 
Mauro Conti \inst{2}
    }
\authorrunning{Dincy et al.} 
\institute{
    Department of Computer Applications, Cochin University of Science and Technology, Kerala, India
    \and
    Department of Mathematics, University of Padua, Padua, Italy
    }

\maketitle              

\begin{abstract}
In the rapidly evolving landscape of cyber security, intelligent chatbots are gaining prominence. Artificial Intelligence, Machine Learning, and Natural Language Processing empower these chatbots to handle user inquiries and deliver threat intelligence. This helps cyber security knowledge readily available to both professionals and the public. Traditional rule-based chatbots often lack flexibility and struggle to adapt to user interactions. In contrast, Large Language Model-based chatbots offer contextually relevant information across multiple domains and adapt to evolving conversational contexts. In this work, we develop \textit{IntellBot}, an advanced cyber security Chatbot built on top of cutting-edge technologies like Large Language Models and Langchain alongside a Retrieval-Augmented Generation model to deliver superior capabilities. This chatbot gathers information from diverse data sources to create a comprehensive knowledge base covering known vulnerabilities, recent cyber attacks, and emerging threats. It delivers tailored responses, serving as a primary hub for cyber security insights. By providing instant access to relevant information and resources, this IntellBot enhances threat intelligence, incident response, and overall security posture, saving time and empowering users with knowledge of cyber security best practices. Moreover, we analyzed the performance of our copilot using a two-stage evaluation strategy. We achieved BERT score above 0.8 by indirect approach and a cosine similarity score ranging from 0.8 to 1, which affirms the accuracy of our copilot. Additionally, we utilized RAGAS to evaluate the RAG model, and all evaluation metrics consistently produced scores above 0.77, highlighting the efficacy of our system. 
\end{abstract}
\section{Introduction} 
\label{sec:introduction}
The rise of interconnected technologies creates new vulnerabilities, leading to a surge in complex cyber threats and demands a faster response to security incidents. So, having some intelligent-based tools like chatbots that can provide threat intelligence about the current landscape is crucial. The chatbots empower cyber security professionals by efficiently answering their queries and keeping them informed about the evolving threat landscape~\cite {lee2023can}. Chatbots are computer programs that mimic human-user conversation through text or voice-based communication~\cite{adamopoulou2020overview}.  
Earlier researchers utilized techniques, such as rule-based or knowledge-based, to implement the chatbot~\cite{ngai2021intelligent}. The rule-based chatbot operates on a set of predefined rules and patterns. However, rule-based systems are not very flexible, and they could have trouble answering complicated or unclear questions and, therefore, do not learn from user interactions~\cite{thorat2020review}. To process user input and provide appropriate responses, chatbots make use of technologies, including Artificial Intelligence~(AI), Machine Learning~(ML), and Natural Language Processing~(NLP).
\par
The emerging landscape of human-computer interaction and NLP has entailed the development of chatbots based on the Large Language Model~(LLM). Unlike traditional rule-based or other chatbot systems, LLM chatbots deliver refined and contextually relevant information to the users. The proficiency of this chatbot extends to tasks such as language comprehension, sentiment analysis, and generating logical responses across various domains~\cite{kim2023chatgpt,ross2023programmer}. Additionally, they offer scalability, effectively handle diverse user queries and adapt to evolving conversational contexts~\cite{birkun2023large}. Recently, security organizations have employed chatbots to explore cyber incidents happening in the real world~\cite{hasal2021chatbots}. A cyber security chatbot is an AI-powered conversational agent designed to assist organizations in addressing cyber security related inquiries, incidents, and concerns~\cite{franco2020secbot}. Security chatbots provide instant support to users, ensuring timely resolution of issues irrespective of temporal constraints. The chatbot frees up valuable human resources, allowing cyber security professionals to focus on more strategic actions and threat mitigation efforts. The chatbot's round-the-clock availability goes beyond the limitations of human support staff, guaranteeing uninterrupted assistance to users worldwide. It helps to reduce the search time for cyber security engineers by providing instant access to relevant information, solutions, and resources, optimizing the troubleshooting and resolution of security incidents.
\par
This paper looks into the architecture, implementation, and advantages of an advanced cyber security chatbot named IntellBot, empowered by cutting-edge technologies such as LLM and Langchain. This chatbot offers cyber security insights and advice, benefiting not only security experts and organizations but also the general public. By providing easily accessible information and resources, it helps users to take proactive measures to protect themselves and their organizations from cyber threats. The efficiency demonstrated by the chatbot results in cost savings by reduced manual intervention and time saved by delivering quick and accurate responses to inquiries. 
\par
The major contributions of this paper include:
\begin{itemize}
\item We compile a security knowledge base from diverse sources, including APT reports, security books, the National Vulnerability Database (NVD), security blog articles, and reports from threat intelligence platforms. Our dataset encompasses a total of 2,447 PDF documents, consisting of 445 cyber security books and 2,002 Advanced Persistent Threat (APT) reports. Additionally, we gathered 7,989 details of malicious file hashes from Virustotal, storing the responses in JSON format. Furthermore, we compile 2,959 URL details. Moreover, we gathered 1,97,256 vulnerability details and scraped 21,825 security blog contents, all saved in CSV format.
    \item We develop an AI-based cyber security copilot by integrating the comprehensive security knowledge base with cutting-edge Large Language Model~(LLM) capabilities.
    \item We evaluate the performance of IntellBot by a two-stage approach. Initially, we analyzed the performance through an indirect approach, in which the BERT score is calculated between the user query and five bot-generated questions from the response to the query. In the second stage, we measure the cosine similarity between the bot-generated responses and human responses for the same query using Word2Vec, Glove, and BERT embeddings. 
\end{itemize}

The rest of the article is organized as follows: Section~\ref{sec:related_work} dives into related research findings, Section~\ref{sec:background} explains the background for the study, Section~\ref{sec:methodology} outlines the methodology employed in this study, Section~\ref{sec:result} showcases the experimental results, including the interface and IntellBot output, Section~\ref{sec:conclusion} concludes the work and proposing future research directions. 
\section{Related work}
\label{sec:related_work}
Large language models excel in various NLP tasks, including creative text generation, translation, and question answering~\cite{yang2023harnessing}. These AI models, empowered by massive amounts of text data, offer versatility due to their ability to handle unseen data and require less task-specific training than traditional models. Notably, LLMs are advantageous when data scarcity is a concern. However, fine-tuned models may still be preferable for tasks with abundant data. While LLMs hold significant promise, real-world applicability, interpretability, and potential biases remain challenges. A few research works are available in the state of the art, integrating NLP and LLM for chatbot creation.
\par
 In~\cite{androutsopoulou2019transforming}, the authors proposed an approach for using chatbots to improve communication between citizens and government. The chatbots help to understand citizens' queries and provide them with information or complete transactions. Pandya et al.~\cite{pandya2023automating} 
 introduced Sahaay, an open-source system that uses LLM to automate customer service. Sahaay directly retrieves information from a company's website and leverages Google's Flan T5 LLM to answer customer queries. The model is further trained with Hugging Face Instructor Embedding to improve its understanding of meaning. Looking to address the limitations of traditional, inflexible methods in entrepreneurship education, a new study ~\cite{ilagan2023prototype} proposed a groundbreaking solution: an LLM chatbot acting as a startup coach simulator. This system tackles the iterative nature of startup development by leveraging a conversational interface powered by LLMs like GPT-3. They created the chatbot to provide real-time feedback on crucial aspects like their business model, product-market fit, and financial projections. In~\cite{mansurova2023development}, authors proposed an LLM-based chatbot for the Fintech industry. They integrated technologies, including Langchain, to enhance response accuracy and relevance. Initial findings suggest a positive impact on user knowledge regarding the Digital Tenge project.

 Oliveira et al.~\cite{oliveira2023introducing} proposed a solution to improve student communication and information access. Their virtual assistant chatbot offers quick and accurate answers frequently asked questions about admissions, programs, registrations, financial aid, and campus resources. This eliminates the need for students to spend time searching for information manually. Building on LLM advancements, Arora et al.~\cite{arora2023llms} introduced JEEBENCH, a challenging dataset assessing reasoning through complex physics, chemistry, and maths problems. They gathered data from engineering exams and identified shortcomings in existing LLMs, such as GPT-4, particularly in maths core concepts, algebra, and arithmetic. This highlights the need for future research on integrating mathematical logic into LLMs and evaluating their decision-making abilities.
\par
A recent study by McIntosh et al.~\cite{mcintosh2023harnessing} introduces a novel chatbot assistant specifically designed for cyber security Governance, Risk Management, and Compliance (GRC). This virtual assistant leverages GPT-based models to focus on policy development within the GRC domain, targeting researchers, practitioners, and organizations. Furthermore, they employ Game Theory to assess the effectiveness of policies generated by GPTs. SecBot~\cite{franco2020secbot} is a conversational agent built using ML and NLP techniques to assist non-experts with cyber security decisions. They implemented Dual Intent and Entity Transformer (DIET), which extracts entities from the text using Conditional Random Fields (CRFs). Their knowledge databases store information on cyber security threats and solutions, allowing SecBot to recommend actions to avoid or mitigate problems. In~\cite{shafee2024evaluation}, Bessani et al. investigated how LLMs can be used to improve threat awareness and detection. The study proposes utilizing the chatbots for tasks like binary classification and Named Entity Recognition(NER) to generate Indicators of Compromises (IoCs) that can be used for proactive cyber threat mitigation. To evaluate their effectiveness, the performance of these LLM chatbots is compared against specialized models using real-world data collected from Twitter. The findings of this research emphasize the value of LLMs in the realm of Cyber Threat Intelligence (CTI) and showcase the project's utilization of GPT-based and open-source chatbot technologies.
\section{Background}
\label{sec:background}
\subsection{Langchain}
Langchain\footnote{https://www.langchain.com/} is a Python library that has been developed to simplify the creation of applications that utilize LLMs. It has emerged as a valuable resource for constructing intelligent chatbots.
The core of Langchain is built on a modular and extensible framework that abstracts the complexities of working with LLMs. It provides a range of fundamental tools for loading and managing LLM models from different sources, tokenizing input text, batching requests, and managing other common LLM operations. A notable feature of Langchain is its compatibility with a diverse selection of LLM models from prominent providers like OpenAI\footnote{https://openai.com/}, Anthropic\footnote{https://www.anthropic.com/}, Cohere\footnote{https://cohere.com/}, and AI21\footnote{https://www.ai21.com/}, as well as open-source models such as LlamaCP\footnote{https://github.com/ggerganov/llama.cpp} and GPT-J\footnote{https://www.eleuther.ai/artifacts/gpt-j}.
\par
Furthermore, Langchain offers advanced tools for enhancing the capabilities of LLMs, which are particularly beneficial in the context of a cyber security chatbot. These tools encompass memory components for preserving conversational context, agents for breaking down complex tasks into manageable subtasks, and mechanisms for integrating knowledge bases.
In building our cyber security chatbot, we leveraged several key components from the Langchain framework. 
These key components are as follows:
\begin{itemize}
    \item \textit{Prompts:} Prompts are crafted instructions given to a language model to guide its responses in specific contexts. Prompts could include structured queries or instructions to ensure that the model provides relevant and accurate information.
    \item \textit{Document Loader:} Document loaders are tools or scripts used to ingest and preprocess documents. This helps to load documents from a file system, database, or any other storage medium and preparing them for input into the language model.
    \item \textit{Agents:} An agent serves as the intermediary between humans and the LLM. It interacts with the LLM to execute tasks or generate outputs according to provided instructions. This can take the form of a script, program, or user-friendly chat interface. The agent receives inputs, crafts a prompt to guide the LLM, sends it to the LLM, and handles the LLM's response to achieve the intended result.
    \item \textit{Chains:} The chain represents the sequence of operations or steps that the agent follows to interact with the language model. This could include preprocessing inputs, sending queries to the language model, receiving responses, and post-processing or formatting the outputs as needed.
    \item \textit{LLM:} LangChain's core engine is the LLM, acting as the language model itself. It is responsible for understanding inputs (prompts or queries) in natural language and generating appropriate responses or outputs based on its training data and algorithms.
\end{itemize}
\subsection{RAG Model}
Understanding how Retrieval-Augmented Generation (RAG) models work with LLM is essential for optimizing chatbots. LLMs lay the groundwork, empowering chatbots to understand and generate human-like text. When integrated with RAG models, which combine retrieval-based and generation-based approaches, chatbots can autonomously derive precise responses from documents. The typical RAG framework begins when a user submits a query or question, typically as a text prompt seeking a detailed and accurate response. This query is initially processed by the retriever model, which searches a large database or document corpus to identify and retrieve the most relevant information. The model employs techniques such as sparse or dense vector search, utilizing methods like TF-IDF, BM25, or dense embeddings from transformer models like BERT. It selects a set of top k relevant documents or passages most likely to contain the necessary information.

Once the relevant documents are retrieved, they are prepared for the generative phase. This preparation involves extracting and formatting the pertinent content to provide context for the generative model. The generative model, often a transformer-based language model like GPT, receives both the original query and the retrieved documents as input. By integrating this information, the model generates a coherent and contextually appropriate response, combining insights from the retrieved documents with its pre-trained knowledge.
This response is then delivered to the user, grounded in the retrieved information, which enhances its reliability and reduces the likelihood of factual inaccuracies or hallucinations. RAG ensures that responses generated by an LLM are not solely dependent on static or outdated training data. Rather, the model utilizes the dataset containing different documents to deliver accurate responses. 

\section{Methodology}
\label{sec:methodology}
\begin{figure}[h!]
    \centering
    \includegraphics[width=0.9\linewidth]{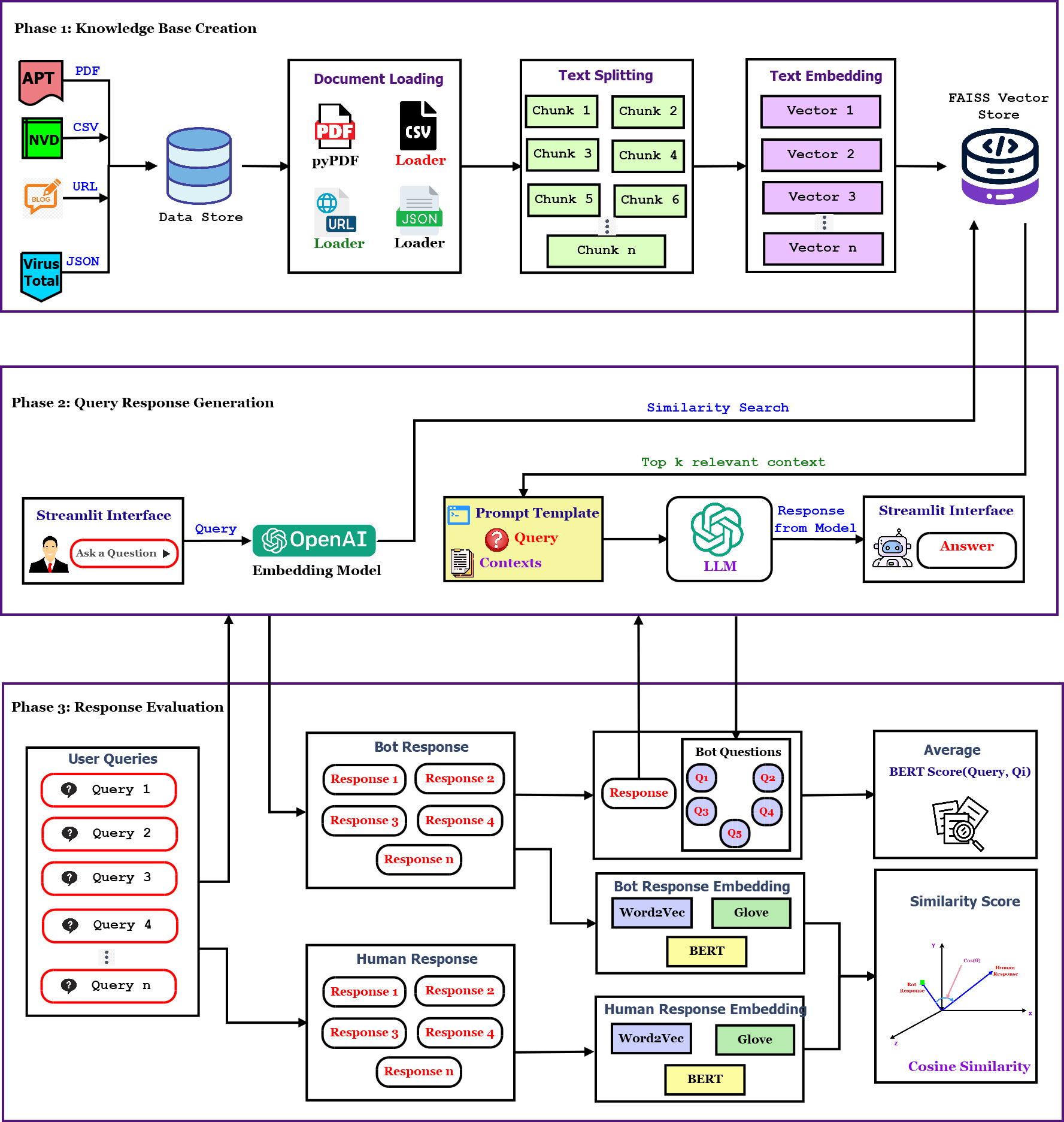}
    \caption{Proposed Architecture of Security chatbot created using the Large Language Model}
    \label{fig:architecture}
\end{figure}

The section outlines the systematic procedure employed in the development of a cyber security chatbot using LLM technology. The comprehensive architecture of IntellBot is depicted in Fig.~\ref{fig:architecture}. Our system encompasses three phases: (i) Creation of the Security Knowledge Base, (ii) Generation of Query Response Interface, and (iii) Evaluation of Chatbot Responses with those generated by humans (Subject Matter Experts). The creation of the security knowledge base involves data collection, document loading, text segmentation, conversion of textual data into numerical format, and storage of embedding vectors in the Facebook AI Similarity Search (FAISS) vector store. Subsequently, a user-friendly interface was developed using Streamlit, enabling users to pose security-related queries. The interface and response generation were implemented utilizing the Large Language Model. Finally, we evaluate the IntellBot's responses using a two-stage strategy involving indirect proof and cosine similarity. Detailed descriptions of all procedures are provided in the subsequent section.

\subsection{Data Collection}
The study employed four different types of documents to create a comprehensive dataset for analysis, which included PDFs, CSVs, JSONs, and URLs. The dataset consisted of a total of $2,447$ PDF documents comprising 
$445$ cyber security books refereed from different GitHub pages and $2,002$ Advanced Persistent Threat (APT) reports sourced from vx-underground\footnote{https://vx-underground.org/APTs}, providing detailed information on various malware attacks and procedures. Also, we collected data from the threat intelligence platform known as VirusTotal\footnote{https://www.virustotal.com/gui/home/upload}. We searched $7,989$ malicious file hashes in Virustotal and saved the response in JSON format.
 Moreover, custom crawlers were developed to collect information from websites such as KrebsOnSecurity and Malwarebytes, focusing on URLs related to cyber threats and security incidents. Additionally, vulnerability information was gathered from the National Vulnerability Database~(NVD)\footnote{https://nvd.nist.gov/vuln/data-feeds}. Furthermore, we scraped security-related blogs from platforms like HackerNews\footnote{https://thehackernews.com/}, BleepingComputer\footnote{https://www.bleepingcomputer.com/news/security/}, CISCO\footnote{https://blog.talosintelligence.com/ }, CSO\footnote{https://www.csoonline.com/in/cybercrime/page/ }, etc., and saved the blog content and title into the CSV file. We collected $1,97,256$ vulnerability details and $21,825$ blog content, which are saved in CSV format. This method of gathering diverse data sources enriched the cyber security dataset, making it more comprehensive and valuable for analysis and research. 
 \par
 Furthermore, we performed some data preprocessing steps that aim to improve data quality and usability. In the case of CSV and JSON data, preprocessing typically includes identifying and removing duplicate entries to ensure data integrity and accuracy. Dealing with HTML data obtained through web crawling requires special attention, as it often contains newline and tab characters that can interfere with the analysis. To tackle this issue, preprocessing techniques are applied to eliminate or substitute these characters, thus maintaining the consistency and cleanliness of the data.
\subsection{Document Loading}
Different document loaders were employed to load data in the formats such as PDFs, CSVs, JSONs, and URLs. The PyPDF Loader is used to load PDF data, load a total of 2,447 PDF documents, and extract metadata, page content, and page numbers. 
The JSON Loader handles JSON files obtained from the Virus Total website. JSON Loader effectively extracted data using the \texttt{jq} Python package.
The RecursiveUrlLoader utilized a tailored extractor function relying on Beautifulsoup to traverse webpages starting from root URLs. It then extracted text content from these pages to facilitate subsequent processing. Additionally, data from 38 CSV files was loaded using the CSVLoader. The \textit{load} function helps to get structured data from these files, making them suitable for further analysis in our research. Also, content scraped from 7,889 web pages of hackernews sites was collected and saved as CSV files. 

\subsection{Text splitting}
Text splitting converts the user's input into smaller units, such as words or tokens, allowing for easier data processing and extracting meaningful features. This breakdown enables us to better understand the user's queries or commands and enhance comprehension. In our work, we utilized the \textit{Recursive Character Text Splitter} to carry out this task, incorporating parameters like \textit{chunk size} and \textit{chunk overlap}. The \textit{chunk size} parameter determines the maximum length of individual segments in which a text document is divided. It specifies each chunk's number of characters, words, or tokens. We fixed the chunk size at $1000$ to ensure consistent and manageable segments for analysis. Also, \textit{chunk overlap} indicates the level of shared content between consecutive text segments. A higher overlap suggests more redundancy among chunks, while a lower overlap may lead to more distinct segments. In our study, we choose a chunk overlap value of $50$ to balance overlap and coherence, maintaining continuity across segmented text data.
\subsection{Conversion of text into Numerical format}

Using embedding techniques is crucial in transforming textual data into a numerical framework. This methodology involves representing chunks of text as vectors, which allows for the effective capture of semantic relationships and similarities. In our work, we employed the \textit{Sentence Transformer model} from the Hugging Face framework to generate these embeddings, specifically employing the model ``sentence-transformers/all-mpnet-base-v2"\footnote{https://huggingface.co/sentence-transformers/all-mpnet-base-v2}. 
Moreover, we leveraged the use of embeddings twice within our chatbot system. Firstly, we utilized embeddings for encoding the text chunks retrieved from the text splitting phase. 
Additionally, we employed embeddings to encode the questions posed by users. To accomplish this, we utilized the OpenAI model (GPT-3.5-turbo), using its capabilities to convert user queries into vector representations. By incorporating embeddings in both stages of the chatbot, we improved the system's capacity to understand and respond to user queries efficiently. This resulted in seamless interaction and enhanced information retrieval within the chatbot system.

\subsection{Creation of Vector Store}
After converting the data into numerical format, we stored it in a vector store, a specialized database designed for efficient storage and retrieval of items.
Each piece of data within the vector store is represented as a high-dimensional vector in a multi-dimensional space, allowing for comparisons and similarity calculations. To enhance efficient storage and query performance, we utilized the Facebook AI Similarity Search (FAISS)\footnote{https://python.langchain.com/docs/integrations/vectorstores/faiss/} framework for constructing and managing the vector store. FAISS is well-known for its capability to optimize similarity search operations, ensuring quick and accurate retrieval of similar items from extensive datasets. We created separate vector stores for each data type, including PDFs, CSV files, URLs, and JSON data. To amalgamate the outputs from these various vector stores and retrievers into a cohesive final response, we employed an \textit{ensemble retriever}. This ensemble method enhances response accuracy and depth by leveraging the unique strengths of each individual retriever. By embracing this approach, we balanced storage efficiency and query performance, enabling swift and effective retrieval of similar items from the vector store. 

\subsubsection{Similarity Search}

Similarity search is a critical component of document retrieval systems, especially in the context of responding to user queries by identifying pertinent documents from a large corpus. 
When a query is submitted, the similarity search algorithm calculates similarity scores between the query and each document, facilitating the retrieval of the top\_k (in our case 3) most similar contexts. These retrieved contexts are then used as the foundation for generating responses to the query. Furthermore, the file names of the retrieved documents are stored in the `metadata' parameter for reference and tracking purposes. After retrieving the most similar documents, the system can create responses to the query by analyzing the content of these documents. By utilizing the information found in the retrieved documents, the system can offer relevant and informative responses that effectively address the user's query. 

\subsection{Designing Prompt}
\label{sec:prompt}
This work uses a prompt-based approach to design a cyber security chatbot. A prompt, in this context, is a structured instruction provided to the chatbot that guides it in formulating a response to user queries. It acts as a bridge between the user’s question and the chatbot’s internal knowledge base. This chatbot is designed to mimic a human cyber security expert by utilizing various information sources, such as context1, context2, and context3, from our vector store and presenting the response. When the user encounters a cyber security issue and poses a query, the chatbot analyzes this context alongside the user’s specific query. 
\begin{center}
\scriptsize
    \small
    \fbox{\parbox{\linewidth}{
    \textbf{[System]}
    
    \textit{"You are a cyber security expert. Provide the responses for the question considering the context below. Your responses should consider factors such as the relevance, accuracy, depth, creativity, and level of detail of their responses."}\\
    
   \textbf{[User Query]}
   
 \{What was the name of the ransomware used by FIN8?\}\\
 
 \textbf{[context 1]}
 
\{the ability to drop arbitrary files and exfiltrate file contents from the compromised machine to an actor-controlled infrastructure. This is not the first time FIN8 has been detected using Sardonic in connection with a ransomware attack...\}\\

 \textbf{[context 2]}
 
\{CONTENT: An exhaustive analysis of FIN7 has unmasked the cybercrime syndicate's organizational hierarchy, alongside unraveling its role as an affiliate for mounting ransomware attacks...\}\\

\textbf{[context 3]}

\{CONTENT: The notorious cybercrime group known as FIN7 has been observed deploying Cl0p (aka Clop) ransomware, marking the threat actor's first ransomware campaign in late 2021...\}}}
\end{center}
\begin{center}
\scriptsize
    \small
    \fbox{\parbox{\linewidth}{
    \textbf{[Response]}
    
    FIN8 has been detected using the White Rabbit ransomware, which is based on Sardonic.}}
\end{center}
\subsection{Streamlit User Interface}

To optimize the interface development for our cyber security chatbot, we used Streamlit\footnote{https://streamlit.io/}, a framework renowned for its simplicity and rapid application development capabilities. Streamlit proved highly advantageous in creating an interactive interface with minimal coding requirements, seamlessly aligning with our data processing framework, LangChain. We used Streamlit version 1.32.2, where it provides a chat\_input function to feed the user queries.
\subsection{Response Evaluation}
\label{sec:res_eval}
After developing the IntellBot for querying security related questions, we need to evaluate the accuracy of its responses. To assess the performance of the bot, we propose a two-stage evaluation approach. Initially, we assessed the correctness of our IntellBot's responses to a query by generating five bot questions based on the retrieved response. We then validated the correctness of the response through indirect proof. Subsequently, we strengthened the evaluation by providing manual answers to the questions and comparing them with the bot's response using similarity scores.

\textit{\textbf{Step 1: Response Evaluation using bot-generated questions}}
\par
In this stage, we employed an indirect proof strategy. We determined the bot answer \( A \) to the cyber security query \( Q_0 \) and generated five subsequent bot questions \( Q_1, Q_2, ..., Q_5 \) based on \( A \). The validation process involved comparing these generated questions with the original query \( Q_0 \) to assess the correctness of \( A \).

Then, we applied indirect proof or proof by contradiction, which asserts that assuming a statement's opposite leads to a logical inconsistency. It relies on the following logic: \textit{If the bot answer \( A \) was incorrect, the questions \( \{Q_i\}_{i=1}^{5} \) derived from \( A \) would logically diverge from \( Q_0 \). However, if the similarity score, \( \text{sim}(Q_0, Q_i) \), is high for all \( i \) (ranges from 1 to 5), it contradicts our initial assumption of \( A \)'s incorrectness. This contradiction suggests that \( A \) correctly addresses \( Q_0 \)}. 

This approach proceeds as follows:
\begin{itemize}
    \item \textit{\textbf{Assumption}}: Initially we assume that bot generated answer \( A \) to the initial query \( Q_0 \) is incorrect.

    \item \textit{\textbf{Implication}}: If \( A \) is incorrect, the subsequent questions \( \{Q_i\}_{i=1}^{5} \) generated by the bot based on \( A \) should not similar to \( Q_0 \).

    \item \textit{\textbf{Proof}}: The process begins with computing the BERT score between \( Q_0 \) and each \( \{Q_i\}_{i=1}^{5} \) to quantify their semantic alignment. BERT score computes similarity scores between pairs of texts using contextual embedding from Bidirectional Encoder Representations from Transformers (BERT). Higher BERT score values indicate stronger semantic coherence between the texts, suggesting a closer alignment in meaning and context. We calculate the average BERT score across the set of questions \( \{Q_i\}_{i=1}^{5} \) generated by \( A \). If this average BERT score yields high values when compared to \( Q_0 \), it implies significant semantic similarity between \( Q_0 \) and the bot-generated questions \( \{Q_i\}_{i=1}^{5} \).

The core of the proof lies in the contradiction analysis: assuming \( A \) is incorrect leads to an expectation that the generated questions \( \{Q_i\}_{i=1}^{5} \) would diverge in meaning from \( Q_0 \). However, if the observed BERT score values between \( Q_0 \) and \( \{Q_i\}_{i=1}^{5} \) are consistently high, it contradicts the initial assumption of \( A \)'s incorrectness. This contradiction indicates that \( A \) likely provides a correct answer to \( Q_0 \).

\end{itemize}
\textit{\textbf{Step 2: Response Evaluation using Human Answer} }

In this step, we further confirm that the IntellBot's response to a query is correct by comparing it with a manual answer. To do this, we curated manual answers to each question before querying IntellBot and collected the chatbot responses via the interface. Subsequently, we compared the bot-generated answer with its corresponding human answer using cosine similarity, which measures how close two sentences are in meaning. To measure cosine similarity, we determine the embeddings of the answers using GloVe \cite{pennington2014glove}, Word2Vec \cite{church2017word2vec}, and BERT. Global Vectors for Word Representation (GloVe) creates dense vector representations of words based on global word co-occurrence statistics, effectively capturing semantic information. Word2Vec generates embeddings by training neural networks on large text corpora, capturing semantic relationships and contextual similarities.

BERT differs from GloVe and Word2Vec in that it is a transformer-based model pre-trained on large-scale text data. BERT generates embeddings that consider both left and right context in all layers, allowing it to capture deep contextual meanings of words and sentences. Finally, we used the Equation~\ref{eq:cosine_similarity} to calculate the similarity between bot and human responses:
\begin{equation}
    CS(A_b, A_h) = \frac{A_b \cdot A_h}{\lVert A_b \rVert \times \lVert A_h \rVert}
    \label{eq:cosine_similarity}
\end{equation}
where $A_b$ and $A_h$ are two vectors representing the bot answer and the human answer, respectively. $\cdot$ represents the dot product operation (sum of products of corresponding elements)
 $\lVert A_b \rVert$ and $\lVert A_h \rVert$ represent the magnitude (length) of vectors $A_b$ and $A_h$, calculated using the L2 norm (square root of the sum of squares of elements).
 
If the similarity score is high, then it indicates that the semantic content of the IntellBot's response is very close to that of a manually curated correct answer. This similarity reflects the accuracy and correctness of the IntellBot's response to a query.
\section{Experimental Result}
\label{sec:result}
\textbf{\textit{Experimental Setup:}} Our experimentation was conducted on a Windows 11 Pro system with an Intel Core i9 processor, 32 GB of RAM, and an NVIDIA Quadro P2000 with 5 GB GDDR5X memory. Additionally, we utilized Google Colab to create the vector store. Our setup utilized various tools and frameworks, including HuggingFace for instructive embedding, RetrievalQA for creating chains, OpenAI for defining the LLM, PromptTemplate for prompt definition, and FAISS for vector storage.
\subsection{Response Evaluation}
In this section, we provide a detailed overview of the query responses received from our chatbot. Also, we evaluate the response using a two-phase approach with bot-generated answers and manual responses.
\par
\begin{table}[h]
\scriptsize
    \centering
    \renewcommand{\arraystretch}{1.7} 
     \caption{BERT Score (BS) comparison between the initial query (\( Q_0 \)) and five bot-generated questions (\( \{Q_i\}_{i=1}^{5} \)).}
    \begin{tabular}{|p{2.5cm}|p{3.2cm}|p{4cm}|p{0.8cm}|p{1.2cm}|}
    \hline
       \textbf{Query}  & \textbf{IntellBot Response} & \textbf{Five Bot Questions} & \textbf{BS} & \textbf{Average BS} \\
       \hline
       \multirow{5}{2.5cm}{Who were the primary victims targeted in the Mirage campaign?} & \multirow{5}{3.2cm}{The primary victims targeted in the Mirage campaign were organizations involved in the contest for rights to survey natural gas and oil in the South China Sea, including a high profile oil company in the Philippines, a military organization in Taiwan, an energy company in Canada, and organizations in Brazil, Israel, Egypt, and Nigeria.} & What is the Mirage campaign and who were the primary targets? & 0.921 & \multirow{5}{1.2cm}{0.904} \\
       \cline{3-4}
         & & How were the organizations involved in the contest for rights to survey natural gas and oil in the South China Sea affected by the Mirage campaign? & 0.878 &  \\
         \cline{3-4}
         & & What were the potential motives behind the Mirage campaign? & 0.924 & \\
         \cline{3-4}
         & & How were the victims identified and targeted in the Mirage campaign? & 0.945 & \\
          \cline{3-4}
          & & What steps can organizations take to protect themselves from similar cyber attacks in the future? & 0.851 &  \\
          \hline
          \multirow{5}{2.5cm}{What infrastructure was used to carry out the Taidoor attacks?} & \multirow{5}{3.2cm}{The Taidoor attacks were carried out using compromised systems, including a primary command-and-control server located in China. The attackers also used various tools and programs, such as an MS-SQL attack tool, CNC programs, and a portable HTTP file server, to carry out their attacks.} & How were the Taidoor attacks carried out? & 0.929 & \multirow{5}{1.2cm}{0.910} \\
       \cline{3-4}
         & & What tools and programs were used by the attackers in the Taidoor attacks? & 0.925 &  \\
         \cline{3-4}
         & & Where was the primary command-and-control server located for the Taidoor attacks? & 0.907 & \\
         \cline{3-4}
         & & How did the attackers gain access to the compromised systems used in the Taidoor attacks? & 0.903 & \\
          \cline{3-4}
          & & What types of attacks were carried out using the MS-SQL attack tool, CNC programs, and portable HTTP file server in the Taidoor attacks? & 0.886 &  \\
         \hline
         
    \end{tabular}
    \label{tab:bert_question}
\end{table}
\begin{table}
\footnotesize
\caption{Cosine similarity score of 10 random questions with its bot answer and human answer. The topics are represented as follows: Vul for vulnerability-related questions, APT for APT attack-related questions, Blog for queries from security blog articles, and VT for VirusTotal reports. Similarity scores were calculated using three different embedding techniques: Word2Vec, GloVe, and BERT, denoted as S1, S2, and S3, respectively.}
    \centering
    \renewcommand{\arraystretch}{1.6} 
    \scriptsize
    \begin{tabular}{|l|p{3cm}|p{3cm}|p{4cm}|l|l|l|}
    \hline
        \textbf{Type} & \textbf{Query} & \textbf{Human Answer} & \textbf{Bot Answer} & \textbf{S1} & \textbf{S2 } & \textbf{S3 }\\ \hline
       Vul &  What versions of the package are vulnerable to Regular Expression Denial of Service (ReDoS) via email validation? & The package forms before 1.2.1, from 1.3.0, and before 1.3.2 are vulnerable to Regular Expression Denial of Service (ReDoS) via email validation. & The package forms before 1.2.1, from 1.3.0 and before 1.3.2 are vulnerable to Regular Expression Denial of Service (ReDoS) via email validation & 0.986 & 1 & 0.998\\
       \hline
        Vul & What vulnerability was identified in SnapGear appliances before the 3.1.4u5 firmware? & denial of service (complete packet loss) via a packet flood & Remote attackers could cause a denial of service by flooding the system with packets. & 0.557 & 0.716 & 0.979\\
        \hline
        APT & What infrastructure was used to carry out the Taidoor attacks? &	C\&C servers and email addresses.	& The attackers used a compromised command-and-control server located in China, as well as other compromised machines that hosted their entire infrastructure. They also used various tools such as MS-SQL attack tool, CNC programs, and FTP servers to carry out the Taidoor attacks. & 0.492 & 0.907 & 0.908\\
        \hline 
        APT & Who were the primary victims targeted in the Mirage campaign?	& The disclosed victims included a high-profile oil company in the Philippines, a military organization in Taiwan, an energy company in Canada, and other entities in Brazil, Israel, Egypt, and Nigeria.	& The primary victims targeted in the Mirage campaign were organizations involved in the contest for rights to
survey natural gas and oil in the South China Sea, including a high profile oil company in the Philippines, a military organization in Taiwan, an energy company in Canada, and organizations in Brazil, Israel, Egypt, and Nigeria. & 0.895 & 0.959 & 0.964\\
\hline
APT & What is the name of the tool used by the Naikon APT group for remote administration of compromised systems? & The tool used by the Naikon APT group for remote administration is called XSControl, which is described as the group's victim management software.	& The tool used by the Naikon APT group for remote administration of compromised systems is called
``xsControl/naikon." & 0.898 & 0.927 & 0.997\\
\hline
  Blog & What type of malware infected the Turkish airline employee's computer and facilitated the theft of Airbus credentials? &	The prevalent and powerful info-stealing trojan called RedLine infected the Turkish airline employee's computer. &	The malware that infected the Turkish airline employee's computer was an info-stealer, specifically the RedLine trojan. & 0.833 & 0.864 &0.996\\
  \hline
Blog & What are some aliases of the Jupyter Infostealer malware? &	Polazert, SolarMarker, and Yellow Cockatoo	& Polazert, SolarMarker, and Yellow Cockatoo. & 0.764 & 1 &0.999\\
\hline
Blog & When did the attack on Kaseya take place?	& The attack on Kaseya occurred over the July 4 weekend in 2021.	& The attack on Kaseya took place on the July 4 weekend in 2021. & 0.946 & 0.943 & 1\\
\hline
VT & List the different versions of Win32/Injector.EDSK? &	26690, 26977, and 26603 & 	The different versions of Win32/Injector.EDSK are 26690, 26977, and 26603. & 1 & 0.701 & 0.977  \\
\hline
VT &	List the different version of Trojan.Win32.PRIVATE LOADER.YXCLPZ? &	10.0.0.1040 and 11.0.0.1006. &	The different versions of Trojan.Win32.PRIVATE LOAD ER.YXCLPZ are ``10.0.0.1040" and ``11.0.0.1006". & 1 & 0.970 & 0.970\\
\hline
           \end{tabular}
    \label{tab:cosine_similarity}
\end{table}

Initially, we developed the IntellBot using the process presented in Section \ref{sec:methodology}. Then we tested various queries with IntellBot related to vulnerabilities, attacks, malware, and campaigns. For instance, one inquiry focused on versions of packages that are vulnerable to Regular Expression Denial of Service (ReDoS) via Email validation, to which the bot accurately responded with package forms before 1.2.1, from 1.3.0, and before 1.3.2. Similarly, we investigated queries related to Taidoor attacks, finding that attackers used a compromised command-and-control server in China, along with other compromised machines, for their infrastructure. They employed tools like MS-SQL attack tools, CNC programs, and FTP servers for these attacks. Subsequently, we evaluated the responses generated by IntellBot and assessed their performance using the two approaches discussed in Section \ref{sec:res_eval}. To do this, we generated 100 questions from each type of document and queried the IntellBot to obtain answers. 

\subsubsection{Response Evaluation Using Bot-generated Questions:}

In this step, we generated five bot questions from each query responses. Then we calculated the BERT score between the initial query (\( Q_0 \)) and each of these five generated questions (\( \{Q_i\}_{i=1}^{5} \)) as described in Section \ref{sec:res_eval}. Subsequently, we determined the average similarity score.

Table \ref{tab:bert_question} presents a sample of questions, their corresponding responses, the five bot-generated questions for the response, their BERT scores, and the average BERT score. From the table, we observed that the BERT score between the initial query (\( Q_0 \)) and the bot-generated questions (\( \{Q_i\}_{i=1}^{5} \)) is consistently very high, nearly 0.91 on average. This indicates that the generated questions closely match the content of the initial query.

We also analyzed all query responses by this method and our analysis reveals that the average BERT score exceeds 0.80 for all \( Q_0 \) and the bot-generated questions \( \{Q_i\}_{i=1}^{5} \). So, based on our indirect proof statement discussed in \ref{sec:res_eval}, the bot-generated response is correct for the specific query. This finding suggests that IntellBot consistently provides accurate and relevant responses to the questions posed, reinforcing its effectiveness in generating appropriate answers.
\subsubsection{Response Evaluation Using Human Answer:}
After evaluating the response using the question comparison approach, we further assessed the response and performance of our IntellBot by comparing it with the manual response. For this, we manually provided an answer for each query and generated a response from IntellBot. To verify the accuracy of the response, we determined the cosine similarity between the bot-generated response and the manual answer. This process involved several text preprocessing steps, including converting text to lowercase, removing stopwords, and splitting sentences into individual words to standardize the input.
\begin{figure}[h!]
\centering
  \includegraphics[width=0.8\textwidth]{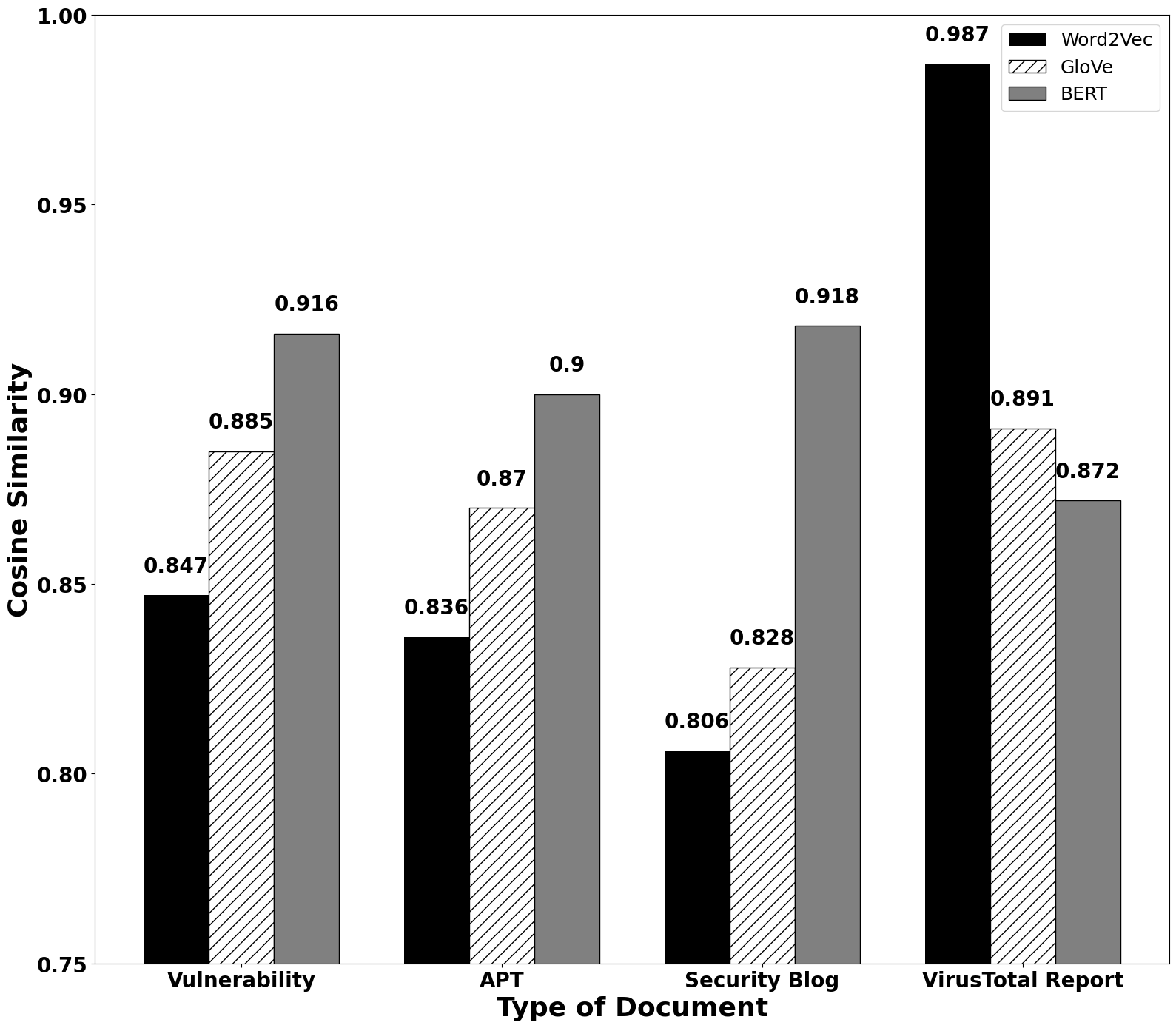}
     \caption{Comparison of average cosine similarity scores generated by Word2Vec, GloVe, and BERT embeddings related to various topics. For each topic, 100 queries were selected, and the cosine similarity between bot and human responses was computed.}\label{Fig:Data1}
\end{figure}
We then converted each word into numerical vectors using GloVe and Word2Vec embeddings, utilizing the \textit{GoogleNews-vectors-negative300.bin} for Word2Vec and \textit{glove.6B.300d.txt} for GloVe to capture semantic meaning. In addition, we calculated cosine similarity using BERT embeddings, which provide a more contextual representation of the text. For this, we initialized the tokenizer and model with \textit{bert-base-uncased}. After computing the embeddings for both the bot responses and the manual answers, we assessed the cosine similarity to quantify the alignment between them.

Cosine similarity between the bot answers and human answers for 10 randomly selected queries, illustrated in Table~\ref{tab:cosine_similarity}. The table reveals that the bot responses closely align with the human responses, showcasing near-identical content. High similarity scores between the bot-generated and human responses affirm the accuracy of our IntellBot-generated answers. Although a few questions exhibit similarity scores below 0.8, we manually verified such questions and found that the bot-generated responses produced correct answers with additional information.
\par
Furthermore, we computed the cosine similarity scores for the responses to 400 queries related to different topics of security. Subsequently, we determined the average similarity score for each type of query. Figure~\ref{Fig:Data1} illustrates the average similarity scores between bot-generated and human responses using different embedding. Across various topics of queries, BERT embedding consistently demonstrates higher similarity compared to Word2Vec and GloVe embeddings. As depicted in Figure~\ref{Fig:Data1}, the similarity scores between bot and human answers range from 0.8 to 1, indicating IntellBot's capability to provide responses closely resembling manual answers in terms of cosine similarity.
\subsection{Evaluation of RAG model}
In this section, we evaluate our RAG model using the Retrieval-Augmented Generation Assessment System (RAGAS) \cite{es2023ragas}. RAGAS provides a comprehensive evaluation by assessing both the retrieval and generation aspects of the model rather than just the final response. For evaluating retrieval systems, RAGAS employs metrics such as \textit{context\_recall} and \textit{context\_precision}. For generation assessment, it utilizes metrics like \textit{faithfulness} to identify hallucinations and \textit{answer\_relevancy} to measure how precisely the answers address the queries. 
\begin{itemize}
    \item \textit{\textbf{Faithfulness:}} Measures the factual accuracy of the generated answer (\( A \)) with respect to the provided context (\( C \)). This is performed in two steps: Given a query (\( Q \)) and the generated answer (\( A \)), an LLM extracts a set of statements \( S(A) = \{s_1, s_2, \ldots, s_n\} \) from \( A \). Then, each statement \( s_i \in S(A) \) is verified against the context \( C \) to determine if it is supported. The faithfulness score \( F \) is computed as:
  \[
  F = \frac{|T|}{|S|}
  \]
  where \( |T| \) is the number of statements supported by the context \( C \) and \( |S| \) is the total number of statements in \( S(A) \).
\item \textit{\textbf{Answer Relevancy:}} Evaluates how relevant and precise the generated answer (\( A \)) is to the query (\( Q \)). RAGAS uses an LLM to generate \( m \) potential questions \( \{q_1, q_2, \ldots, q_m\} \) that the answer \( A \) could address. For each question \( q_i \), embeddings are obtained and compared with the embedding of the original query \( Q \). The answer relevancy score \( AR \) is computed as:
  
  \[
  AR = \frac{1}{m} \sum_{i=1}^{m} \text{sim}(Q, q_i)
  \]
  
  where \( \text{sim}(Q, q_i) \) denotes the cosine similarity between the embeddings of \( Q \) and \( q_i \).
\item \textit{\textbf{Context Recall (CR):}} Assesses the retriever's ability to retrieve all necessary information for answering the query. RAGAS compares the retrieved context (\( C \)) with the ground truth answer (\( G \)) by checking if each statement in \( G \) is present in \( C \). The context recall score \( CR \) is computed as:
  
  \[
  CR = \frac{|G_s in C|}{|G_s|}
  \]
  
  where the \( G_s in C \) is the count of statements from \( G \) that are present in \( C \), and the \( G_s \) is the total number of statements in \( G \).
\item \textbf{\textit{Context Precision (CP):}} Evaluates the accuracy of the retrieved or generated content by measuring the proportion of relevant statements among all the retrieved statements. It is calculated as the ratio of the number of relevant statements retrieved to the total number of statements retrieved.
\[ \text{CP} = \frac{|G_{\text{in}} C|}{|C_s|} \]

where \( |G_{\text{in}} C| \) is the count of statements from the ground truth answer \( G \) that are present in the retrieved context \( C \). \( |C_s| \) is the total number of statements in the retrieved context \( C \).

\end{itemize}
Table \ref{tab:ragas} presents the average scores for four metrics across different types of cyber data sources: Vulnerability, APT Report, Security Blogs, and VirusTotal Report. For Vulnerability-related queries, the system exhibits a high Context Precision of 0.934 and Context Recall of 0.933, indicating accurate and comprehensive retrieval. The Faithfulness score of 0.908 and the Answer Relevancy score of 0.855 further demonstrate that the generated answers are highly accurate and relevant. The queries related to APT Reports show solid performance with a CP of 0.881 and CR of 0.847, but slightly lower compared to vulnerabilities, suggesting minor improvements in retrieval completeness. The Faithfulness score of 0.878 and the Answer Relevancy score of 0.922 are strong, indicating that generated answers are highly relevant and accurate. The queries about information posted in security blog articles have a lower CP of 0.805 but a higher CR of 0.889, showing that while the system retrieves the most relevant information, it also includes some irrelevant data. The Faithfulness score of 0.871 and AR score of 0.883 are good. Also we asked the queries related to VirusTotal Reports of malicious hash and it provides scores near to 0.80 across all metrics reflecting challenges in both retrieval accuracy and answer relevance. Overall, these results indicate that the system performs well across various types of cyber data. The implementation details are available on our GitHub page: \url{https://github.com/OPTIMA-CTI/IntellBot}.
\begin{table}[]
\scriptsize
\caption{Retrieval and Generation metric average score of each type of cyber data. }
    \centering
    \renewcommand{\arraystretch}{1.1} 
    \begin{tabular}{|l|p{1cm}|p{1cm}|p{1cm}|p{1cm}|}
        \hline
        \textbf{Type} & \multicolumn{2}{|c|}{\textbf{Retrieval}} & \multicolumn{2}{|c|}{\textbf{Generation}} \\
        \cline{2-5}
             & \textbf{CP} & \textbf{CR} & \textbf{F} & \textbf{AR} \\
             
        \hline
        Vulnerability & 0.934 & 0.933 & 0.908 & 0.855 \\ \hline
        APT Report & 0.881 & 0.847 & 0.878 & 0.922 \\ \hline
         Security Blogs & 0.805 & 0.889 & 0.871 & 0.883 \\ \hline
        VirusTotal Report & 0.786 & 0.805 & 0.766 & 0.798 \\
        \hline
    \end{tabular}
    \label{tab:ragas}
\end{table}

\section{Conclusion}
\label{sec:conclusion}
The security chatbot, IntellBot, developed in this work demonstrates a novel approach to leveraging Large Language Models and vector databases for answering queries related to cyber security. The IntellBot utilizes the \textit{LangChain} framework, which enables the architecting of Retrieval-Augmented Generation systems with numerous tools to transform, store, search, and retrieve information that refines language model responses. 
Combining LLMs with vector databases highlights the potential for intelligent information retrieval systems tailored to cyber security. We assessed the IntellBot performance using a two-stage approach. Initially, we analyzed performance through an indirect method by calculating the BERT score between the user query and five bot-generated questions from the response to that query. In the second stage, we measured the cosine similarity between the bot-generated responses and human responses for the same query using Word2Vec, GloVe, and BERT embedding. Our findings show that in the first stage, we achieved a BERT score greater than 0.8 and consistently obtained cosine similarity scores ranging from 0.8 to 1, indicating highly accurate query responses. Also, RAGAS's evaluation metrics, consistently above 0.77, highlight the system's effective performance across various types of cyber queries. In the future, we plan to integrate more data sources to enhance the chatbot's capabilities and expand its relevance in cyber security contexts. Moreover, we intend to incorporate more evaluation techniques to comprehensively assess the chatbot's responses. Additionally, we plan to leverage reinforcement learning with human judgment to refine and optimize the bot's responses.
\section*{Acknowledgment}
\begin{figure}[!htb]
\includegraphics[width=3.8cm]{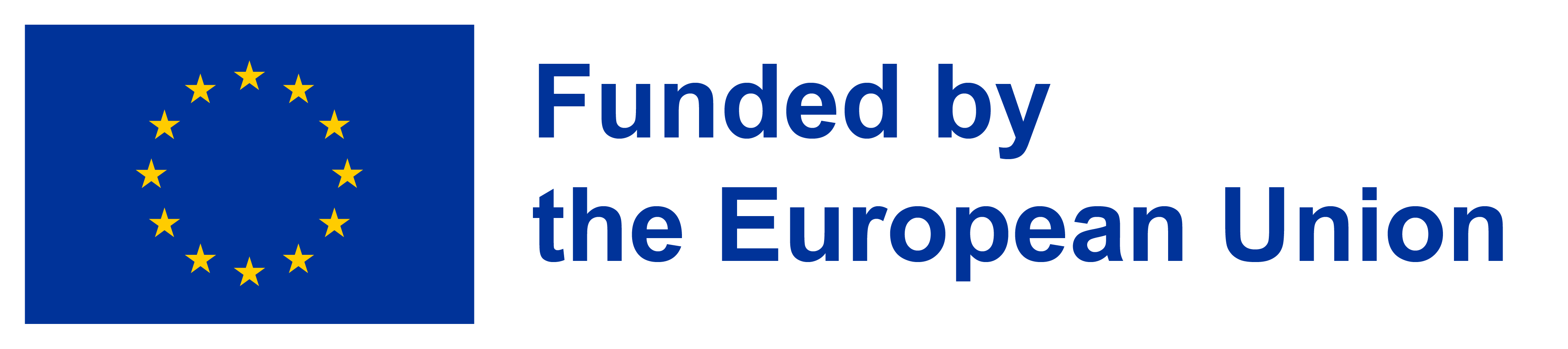}
\end{figure}
This work was partly supported by the HORIZON Europe Framework Programme through the project ``OPTIMA-Organization sPecific Threat Intelligence Mining and sharing"~(101063107), funded by the European Union. Views and opinions expressed are however those of the author(s) only and do not necessarily reflect those of the European Union. Neither the European Union nor the granting authority can be held responsible for them.
\bibliographystyle{unsrt}
\bibliography{sample-base}

\begin{thebibliography}{10}

\bibitem{lee2023can}
Ju~Yoen Lee.
\newblock Can an artificial intelligence chatbot be the author of a scholarly article?
\newblock {\em Journal of educational evaluation for health professions}, 20, 2023.

\bibitem{adamopoulou2020overview}
Eleni Adamopoulou and Lefteris Moussiades.
\newblock An overview of chatbot technology.
\newblock In {\em IFIP international conference on artificial intelligence applications and innovations}, pages 373--383. Springer, 2020.

\bibitem{ngai2021intelligent}
Eric~WT Ngai, Maggie~CM Lee, Mei Luo, Patrick~SL Chan, and Tenglu Liang.
\newblock An intelligent knowledge-based chatbot for customer service.
\newblock {\em Electronic Commerce Research and Applications}, 50:101098, 2021.

\bibitem{thorat2020review}
Sandeep~A Thorat and Vishakha Jadhav.
\newblock A review on implementation issues of rule-based chatbot systems.
\newblock In {\em Proceedings of the international conference on innovative computing \& communications (ICICC)}, 2020.

\bibitem{kim2023chatgpt}
Jin~K Kim, Michael Chua, Mandy Rickard, and Armando Lorenzo.
\newblock Chatgpt and large language model (llm) chatbots: The current state of acceptability and a proposal for guidelines on utilization in academic medicine.
\newblock {\em Journal of Pediatric Urology}, 2023.

\bibitem{ross2023programmer}
Steven~I Ross, Fernando Martinez, Stephanie Houde, Michael Muller, and Justin~D Weisz.
\newblock The programmer’s assistant: Conversational interaction with a large language model for software development.
\newblock In {\em Proceedings of the 28th International Conference on Intelligent User Interfaces}, pages 491--514, 2023.

\bibitem{birkun2023large}
Alexei~A Birkun and Adhish Gautam.
\newblock Large language model-based chatbot as a source of advice on first aid in heart attack.
\newblock {\em Current Problems in Cardiology}, page 102048, 2023.

\bibitem{hasal2021chatbots}
Martin Hasal, Jana Nowakov{\'a}, Khalifa Ahmed~Saghair, Hussam Abdulla, V{\'a}clav Sn{\'a}{\v{s}}el, and Lidia Ogiela.
\newblock Chatbots: Security, privacy, data protection, and social aspects.
\newblock {\em Concurrency and Computation: Practice and Experience}, 33(19):e6426, 2021.

\bibitem{franco2020secbot}
Muriel~Figueredo Franco, Bruno Rodrigues, Eder~John Scheid, Arthur Jacobs, Christian Killer, Lisandro~Zambenedetti Granville, and Burkhard Stiller.
\newblock Secbot: a business-driven conversational agent for cybersecurity planning and management.
\newblock In {\em 2020 16th international conference on network and service management (CNSM)}, pages 1--7, University of Zurich, 2020. IEEE.

\bibitem{yang2023harnessing}
Jingfeng Yang, Hongye Jin, Ruixiang Tang, Xiaotian Han, Qizhang Feng, Haoming Jiang, Shaochen Zhong, Bing Yin, and Xia Hu.
\newblock Harnessing the power of llms in practice: A survey on chatgpt and beyond.
\newblock {\em ACM Transactions on Knowledge Discovery from Data}, pages 1--23, 2023.

\bibitem{androutsopoulou2019transforming}
Aggeliki Androutsopoulou, Nikos Karacapilidis, Euripidis Loukis, and Yannis Charalabidis.
\newblock Transforming the communication between citizens and government through ai-guided chatbots.
\newblock {\em Government information quarterly}, 36(2):358--367, 2019.

\bibitem{pandya2023automating}
Keivalya Pandya and Mehfuza Holia.
\newblock Automating customer service using langchain: Building custom open-source gpt chatbot for organizations, 2023.

\bibitem{ilagan2023prototype}
Joseph~Benjamin Ilagan and Jose~Ramon Ilagan.
\newblock A prototype of a chatbot for evaluating and refining student startup ideas using a large language model.
\newblock {\em EdArXiv Preprints}, 05 2023.

\bibitem{mansurova2023development}
Aigerim Mansurova, Aliya Nugumanova, and Zhansaya Makhambetova.
\newblock Development of a question answering chatbot for blockchain domain.
\newblock {\em Scientific Journal of Astana IT University}, pages 27--40, 2023.

\bibitem{oliveira2023introducing}
Pedro~Filipe Oliveira and Paulo Matos.
\newblock Introducing a chatbot to the web portal of a higher education institution to enhance student interaction.
\newblock {\em Engineering Proceedings}, 56(1):128, 2023.

\bibitem{arora2023llms}
Daman Arora, Himanshu~Gaurav Singh, and Mausam.
\newblock Have llms advanced enough? a challenging problem solving benchmark for large language models, 2023.

\bibitem{mcintosh2023harnessing}
Timothy McIntosh, Tong Liu, Teo Susnjak, Hooman Alavizadeh, Alex Ng, Raza Nowrozy, and Paul Watters.
\newblock Harnessing gpt-4 for generation of cybersecurity grc policies: A focus on ransomware attack mitigation.
\newblock {\em Computers \& security}, 134:103424, 2023.

\bibitem{shafee2024evaluation}
Samaneh Shafee, Alysson Bessani, and Pedro~M Ferreira.
\newblock Evaluation of llm chatbots for osint-based cyberthreat awareness.
\newblock {\em arXiv preprint arXiv:2401.15127}, 2024.

\bibitem{pennington2014glove}
Jeffrey Pennington, Richard Socher, and Christopher~D Manning.
\newblock Glove: Global vectors for word representation.
\newblock In {\em Proceedings of the 2014 conference on empirical methods in natural language processing (EMNLP)}, pages 1532--1543, 2014.

\bibitem{church2017word2vec}
Kenneth~Ward Church.
\newblock Word2vec.
\newblock {\em Natural Language Engineering}, 23(1):155--162, 2017.

\bibitem{es2023ragas}
Shahul Es, Jithin James, Luis Espinosa-Anke, and Steven Schockaert.
\newblock Ragas: Automated evaluation of retrieval augmented generation.
\newblock {\em arXiv preprint arXiv:2309.15217}, 2023.

\end{thebibliography}
\end{document}